# Uncertainty estimation on frequency shift of Brillouin light scattering

Patrice Salzenstein [1,*]

[1] Centre National de la Recherche Scientifique (CNRS), Franche-Comté Electronique Mécanique Thermique Optique Sciences et Technologies (FEMTO-ST) Institute, Université de Franche-Comté (UFC), Besançon, France

* Correspondence: patrice.salzenstein@cnrs.fr

**Abstract:** Better knowing the precision on the measured value of the Brillouin peak frequencies is essential in order to use it to deduce parameters related to studied materials. Modern methods for evaluating uncertainties are based on the recommendations of the Guide to the expression of uncertainty in measurement. After checking the agreement between the measured 15.7 GHz shift on a Brillouin signal on Poly-(methyl methacrylate) and the expected value, the elementary uncertainty terms are evaluated in two groups, by statistical methods and by other means. We describe the general principle of operation of Brillouin light scattering setup with a high power laser and evaluate the elementary terms of uncertainty, according to international standards enacted for metrology. The global uncertainty on frequency shift is then calculated and we find $\pm 1.20 \times 10^{-2}$ at 2 $\sigma$.

## 1. Introduction

Brillouin light scattering (BLS) is gradually gaining popularity in various industrial applications and in laboratories. This work is about BLS and focus on the discussion about the associated uncertainty.

BLS is the inelastic scattering of light through sound waves. BLS is then a good way to study the elastic properties of materials. It is a non-contact, non-destructive method, and relatively easy to implement with appropriate means. We propose here to come back to the main steps, which allow having specific instrumentation in order to estimate the speed of propagation of phononic waves in materials. Of course, we realize that there are necessary aspects of theoretical knowledge. It is not a question here of going back to all the theories in order to understand the matter. This would require going too deeply in a systematical description. To better frame the subject that we are discussing here, we remind about the main contributions to detect sound waves, by brothers Curie [1], and inelastic light scattering in materials thanks to their excitation by sound waves with Brillouin [2]. The whole instrumental aspect has importance as modern benches for Brillouin diffusion need the appropriate technology. Among other necessities, there are interferometry technologies. In addition, such a Brillouin scattering bench needs conventional optical components, and primarily a sufficiently powerful laser.

Stimulated Brillouin scattering is a nonlinear process. It can also occur in optical fibers. It manifests itself concretely through the creation of a backward propagating Stokes wave, which carries most of the input power, once the one reaches Brillouin threshold [3]. The magnetic properties of materials are via their magnetic excitations - magnons, thanks to Brillouin scattering. As with phonons, magnons concern surface and bulk excitations [4]. Indeed, Brillouin inelastic light scattering spectroscopy is widely used for the study of phonons but also magnons in materials. This technique has become an essential tool [5, 6]. It is of course complementary to inelastic light scattering Raman spectroscopy [7]. Kojima shows that those techniques have become essential for studying materials science [8]. Grimsditch and Ramdas have made precise measurements with Brillouin scattering in the early seventies on Diamond [9]. It is also useful to recall the differences between Brillouin scattering and another well-known technique, Raman spectroscopy. The latter type of spectroscopy is used to determine the chemical composition and molecular structure of the transmission medium, while Brillouin scattering can be used to measure the elastic behavior of a material. A more systematic method was implemented based on a tandem Fabry–Perot interferometer by Sandercock

[10] in 1970 and then by Lindsay, Anderson and Sandercock [11], and Dil et al [12]. Hillebrands [13] and Scarponi et al [14] improved corresponding instrumentation techniques.

Our objective is to assess with a consistent metrological approach the uncertainties of BLS. This paper provides an experimental part and the determination of the uncertainties associated with the determination of peaks corresponding to the shift between the frequency of the signal refracted by a material and to the laser serving as an interrogation signal. The knowledge of this value, as well as the parameters of the studied materials, can also provide the value of the speed of the corresponding phononic waves, when intrinsic characteristic data of the evaluated materials are known. To lead the discussion on the uncertainty associated with the BLS, we rely on the standards of metrology.

The work is organized as follows: section 2 presents materials and methods, section 3, results and verification. Section 4 consists in a discussion about the uncertainty.

## 2. Materials and Methods

The analysis mainly consists in a method of detection of the refracted light emitted by a material under test, i. e. the Device under Test (DUT). This material can be isotropic or anisotropic. One of the key of the measure is the Tandem Fabry-Perot double interferometer. Detected peaks are shifted from the wavelength of the laser. Those offset frequencies depend on the properties of the material of the DUT. This paper aims to lead an estimation of the uncertainty obtained on the frequency shift that can lead later to parameters of the material like phase velocity of transverse and longitudinal waves, deduced from BLS. Estimating the uncertainty requires knowing the contribution of the different fixed parameters like the optical index, the wavelength, the diffusion angle, the density of the material, and the longitudinal and shear modulus, but especially fluctuation of the source, mechanical stability of the setup, and environmental parameters in the room. For this uncertainty estimation, we use a similar method like in optics and microwaves based on the requirement delivered by the *Bureau International des Poids et Mesures* (BIPM) in the guide "Evaluation of measurement data – Guide to the expression of uncertainty in measurement (GUM)" [15].

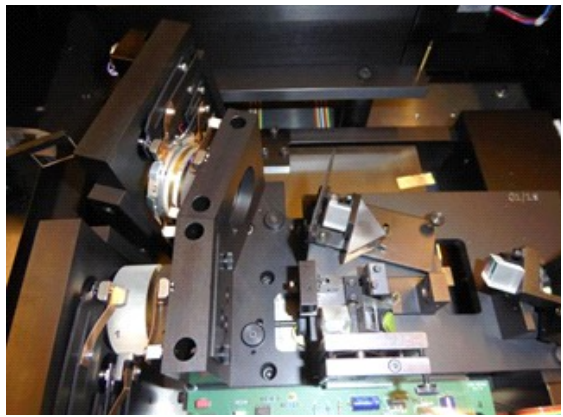

**Figure 1.** Photography of the double Fabry-Pérot interferometer.

We will focus in this part on the principle of Brillouin light scattering. BLS using a 532 nm powerful Class 4 laser up to 600 mW is efficient to reveal spin wave or acoustic signals, at frequencies from few Giga Hertz to more than a hundred of Giga Hertz. Fluctuations of refractive index in a medium enables the detection and analysis of laser light scattered, thanks to BLS setup [11, 12]. The double interferometer is visible on Figure 1. The general principle is to send the signal generated by the laser focusing it on the part of the sample that we want to characterize. The photons arrive in the material or in the thin layer and interact with the lattice or more generally with the material.

Light helps to create phonons. These phonons propagate with speeds that may be different depending on whether the mode is transverse or longitudinal. It depends on the nature of the material, as it can be isotropic or anisotropic. The phonons in turn create light, which are shifted in frequency relatively to the wavelength of the laser. The BLS precisely consists in analysis of the refracted light emitted by a material [11, 13].

Tandem Fabry–Perot interferometer produces peaks shifted from the frequency of the laser to characteristic frequencies depending on the material. Figure 2 gives the typical set-up used for the measurement, showing the typical setup (a) and a picture of the system (b).

We calibrated the bench with part of the laser signal, used as the bench reference. Inside the commercial bench developed by the Swiss company "The time Stable", the light goes with six passages through two different interferometers. Each pair of mirrors is very precisely aligned during the calibration procedure.

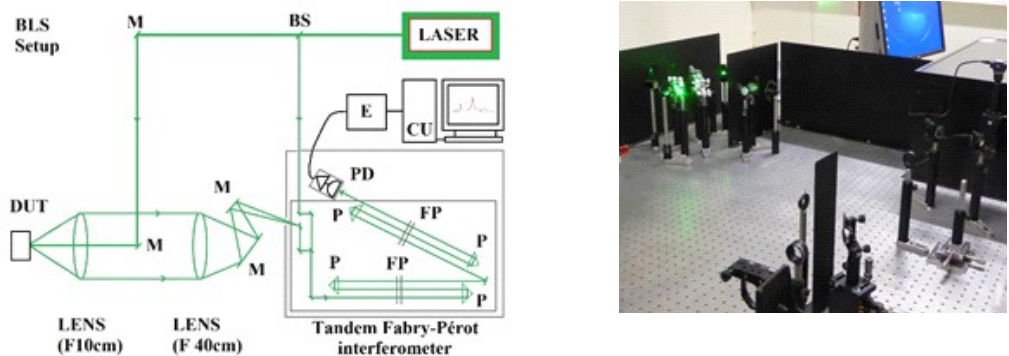

**Figure 2. (a):** Typical setup for BLS. JRS TFP2 is a commercial Tandem Fabry-Pérot interferometer. BLS: Brillouin Light Scattering. DUT: device under test. M: mirror. FP: Fabry-Pérot. P: prism. PD: photodetector. E: electronics. CU: computer unit. **(b):** The commercial Tandem Fabry-Pérot interferometer is inside the box on the right side of this picture.

It is necessary to calibrate accurately the instrument. It is sensitive to mechanical vibrations, temperature and hygrometry. Alignment process requires an alignment of the two cavities. Each of the two cavities consists in a pair of parallel mirrors. Tandem interferometer produces two series of absorption peaks with respect to a flat noisy intensity level. We then obtain a curve providing the number of absorbed photons versus frequency.

We detail in section 4 the different contributions of some of those elements to the global uncertainty on the measured signal.

## 3. Results and verification

Experiments reproduce known Brillouin scattering peaks of bulk materials and some thin films.

Typical Brillouin scattering stimulation reveals acoustic or spin waves frequencies in the range between 3 and 150 GHz, although this is generally limited to around thirty Giga Hertz. In this section, we provide some examples of spectra with detected photons versus frequency shift for isotropic or anisotropic materials.

As a way of calibration, we can characterized a bulk material such as for example Poly-(methyl methacrylate) (PMMA). This isotropic material exhibit well-defined peaks.

From BLS we can deduce parameters of the material like phase velocity of transverse and longitudinal waves: Knowing *n* (optical index), λ (wavelength), θ (diffusion angle), v (phase velocity of transverse or longitudinal waves), *ρ* (density of the material), $c_{11}$ and $c_{44}$ (longitudinal and shear modulus), the Brillouin frequency $\nu_B$ is given by (i):

$$\nu_B = \frac{2nv}{\lambda_0} \sin\left(\frac{\theta}{2}\right)$$

As a concrete example of an isotropic material, we check PMMA, given by the curve on figure 3.c. with λ=532nm, n=1.49, $c_{11}$=9 GPa, $\rho$=1.19x$10^3$ kg/$m^3$, θ=180°, $v_L$=($c_{11}/\rho$)$^{-1/2}$=2750 m/s, we can check that $\nu_B$=15.7 GHz. We demonstrated a good agreement between predicted value for frequency peak and its measured value.

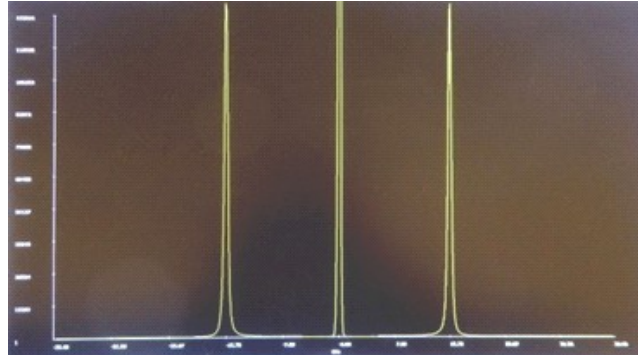

**Figure 3.** BLS spectrum for PMMA with peaks at 15.7 GHz. Frequency shift expressed in Giga Hertz is on horizontal axis. Vertical axis corresponds to the number of detected photons; it can be in an arbitrary unit.

This curve of measurements, given in Fig. 3, is primarily for illustrative purposes. We will focus in the next section on how we can trust the measured frequency values.

Note that for a material like sapphire, which is isotropic, the peaks will depend on the orientation of the DUT sample to be measured. In this case, it would be useful to take an interest in the curves of slowness in the space of the k-vectors, corresponding to the orientation of the sample with respect to the laser signal sent to such DUT.

**4. Discussion about the uncertainty**

In this section, we aim to lead an estimation of the uncertainty on the frequency shift induced by BLS.

Before going into more details in how we may estimate the uncertainty, it is useful to think about the approach in its determination. In the scientific community, it is important to underline that a debate exists as to whether there is a true value. Thomas von Clarmann et al offer the benefit of a critical discussion on the error concept versus the uncertainty concept [16]. Jong Wha Lee et al [17] compare the realist view of measurement and uncertainty versus the instrumentalist view of measurement when quantities are not natural attributes of the world that exist independently of the human perception. They show that a clear understanding of the two views is critical for understanding the guide "Evaluation of measurement data – Guide to the expression of uncertainty in measurement (GUM)" [15].

Estimating the uncertainty requires the knowledge of the contribution of the different fixed parameters, such as the optical index, the wavelength, the diffusion angle, the density of the material, and the longitudinal and shear modulus, but especially fluctuation of the source, mechanical stability of the setup, and environmental parameters in the room. From the equation given in the previous part, we see that the phase velocity of the transversal or longitudinal waves linearly depends on $\nu_B$ (the Brillouin frequency), n (the optical index), λ (the wavelength), Q (the diffusion angle).

The estimation of uncertainty follow the modern way of performing it [18]. For this uncertainty estimation, we use a similar method like in optics [19 – 21] and microwaves [22, 23] based on the requirement delivered by the Bureau International des Poids et Mesures (BIPM) in its guide "Evaluation of measurement data – Guide to the expression of uncertainty in measurement (GUM)", available in reference [15].

The uncertainty in the result of a measurement consists of several components, which may be easily grouped into two main categories according to the way in which their numerical value is estimated.

4.1. Statistical contributions

Following the guidelines, the first category is called A–type. It corresponds to contributions evaluated by statistical methods such as reproducibility, repeatability. We can point out here the variations of results versus time, or with various operators. There is also a question of finding a good compromise between measuring fast enough from few minutes to few hours, and increasing the resolution by having enough samples, as we choose to have at least 2048 samples.

Repeatability $A_1$ is the variation in measurements obtained by one person on the same item and under the same conditions. Repeatability conditions include the same measurement procedure, the same observer, the same measuring instrument, used under the same conditions, repetition over a short period of time, the same location. As the operator does not change, this term is chosen to be zero.

For the reproducibility $A_2$, the same operator perform the measurements. There is no changes caused by differences in operator behavior. All components and devices are dedicated to the instrument, and none of them is replaced. This term is chosen to be zero.

Frequency resolution $A_3$ of the measurement depends on the number of samples, i. e. the difference between two measurement frequency points along the horizontal axis. For 2048 points on a 30 GHz scale, we have a 15 MHz interval. The characteristic peak of Brillouin is a Lorentzian distribution [24], also known as a Cauchy distribution, which is a probability density function. Lorentzian, noted L(f) versus the frequency shift of the optical signal is given in the following expression (ii):

$$L(f) = \left(\frac{\gamma}{\pi}\right) * \left[\frac{1}{\gamma^2+(f-f0)^2}\right] = \left(\frac{1}{\pi\gamma}\right) * \frac{1}{1+(f-f0)^2/\gamma^2}$$

(ii)

where γ is half of the width at half maximum (FWHM): γ=FWHM/2, $f_0$ is the true value of the Brillouin frequency peak. For a given true value, with a peak of great smoothness, we will have in the worst configuration 3 points, which will allow us to approximate a curve in the form of Lorentzian. BLS on PMMA, Nylon or glass show well-defined peaks for isotropic materials. Considering a rectangular distribution, we deduce that $A_3$=0.0005/√3=2.89x10$^{-4}$.

Finally, statistical contribution is:

(iii)  $A = \sqrt{(\Sigma A_i^2)}$

We assume that it is not negligible, even when we track peaks up to 150 GHz from the optical carrier at a 532 nm wavelength, regarding the contributions of other elementary terms we describe in the next sub-section. The more samples there are, the higher the peak is. However, it has no influence on the precision of the frequency shift. Concerning the contribution of the elementary terms of statistical type, we can see the impact of the resolution on the uncertainty of a measured peak, but there is still a risk of not detecting a peak if the sampling frequency is too low.

4.2. Contributions evaluated by other mean

Second family of uncertainties contributions is for those assessed by other means. They are called B–type and depend on various components and temperature control. Experience with or general knowledge of the behavior and properties of relevant materials and instruments determines them.

Frequency references of the 5 MHz or 10 MHz type possibly ensures the traceability of the BLS method to national standards [25, 26]. Indeed, it is then possible to have the best references in terms of frequency stability to connect them to additional measuring devices such as oscilloscopes or any means of frequency measurement.

When there is no traceability or calibration certificates, we refer to manufacturer's specifications, data provided in calibration and other certificates, or uncertainties assigned to reference data taken from handbooks. Such terms are called $B_R$. BLS is not referenced to a standard, as the method is intrinsic. So the data provided in calibration and other certificates, noted $B_R$, are not applicable. It results that we can keep zero as a good approximation of BR.

Other elementary terms are noted BL: the components in B–type category should be characterized by quantities, which may be considered as approximations to the corresponding variances.

We find here the contribution of the laser, noted $BL_1$, as it mainly contributes with its uncertainty on the wavelength, but also with its uncertainty relative to the beam diameter given at 1.7 ±0.2 mm, according to the manufacturer Laser quantum Ltd., to the pointing stability less than 2 µrad/°C, and the beam angle less than 1 mrad. It may contribute to geometrical error, especially in the double interferometer. Some cosine error can then occur. The laser beam and the axis of displacement are not completely parallel [27, 28]. If we call A the angle between the two axis (beam axis and displacement axis) we have then an elementary term of error $e_A$ = L(cosA – 1), which is approximatively equal to – LxA²/2 as A<<1. For a 1 mm distance, we have A up to $10^{-4}$ and $e_A$ up to $8x10^{-11}$. It leads only to an is negligible elementary term. Another contribution to this term is called Abbe error [29, 30]. It corresponds to the magnification of angular error over distance. An angular error of 1 degree corresponds to a positional error of over 1.745 cm, equivalent to a distance-measurement error of 1.745%. In our case, we have an error up to 0.000085%. This elementary term of error of Abbe is dominating on the error of parallelism. Assuming a rectangular distribution of this term, we deduce $BL_1$=4.91x$10^{-7}$.

Contribution of the laser to the noise is noted $BL_2$. Relative Intensity Noise (RIN) of lasers are related to the ratio between the average of the square of the fluctuation optical power (δφ) on the the square of the average optical power $φ_0^2$.

(iv) $\qquad$ RIN(ω)=<│δφ│²> / $φ_0^2$

where ω is the pulsation. RIN generally present a floor until the Fourier frequency, which is equal to the relaxation frequency of the laser. Then the noise decreases. This relaxation frequency is generally in the range of one Mega Hertz. Datasheet of the Torus 532 nm laser (Laser quantum Ltd.) indicates a RIN no worse than -135 dB/Hz at 10 GHz. Using a Fabry-Perot interferometer (JRS Scientific Instruments) the torus laser typically shows high spectral purity with side bands <-110 dB compared with the central mode. This laser is set to operate in normal conditions between 15 to 35ºC. We can consider that $BL_2$=5.77x$10^{-11}$.

PD contribution is $BL_3$. Datasheet of Hamamatsu H10682-210 indicates specifications in the range of -20 to +50°C, with count sensitivity respectively at wavelengths of 500 nm and 600 nm typically between 4.6 x $10^5$ and 1.3 x $10^5$ $s^{-1}.pW^{-1}$. We assume it is not limitative for photons detected during measurements. This contribution is negligible for the effect on the frequency shift.

We consider the contribution of temperature, noted $BL_4$. The temperature varying less than 0.1°C during measurement, despite it can vary from 0.5 degrees during a whole day. Temperature variation in the laboratory is in the range 21 – 25 °C. The maximum variation is ±2°C. It is compatible with what is written in the datasheet of the double interferometer : "The temperature of the environment should be regulated to better than ±2°C over a 24 hour period." Its influence in terms of noise is $e_{Temp}$=10xLog (298/296) =0.0292dB. This distribution is rectangular. It is important to clarify that these variations are slow variations, and that the double interferometer is stable against temperature changes. We deduce that $BL_4$=3.89x$10^{-3}$.

Contribution on the value of the wavelength relies on environmental conditions such as pressure and humidity. We call it $BL_5$. Under normal laboratory measurement conditions, the contribution of small pressure variations and of relative humidity remain negligible. The measurements do not show any dependence to those environmental contributions. This term $BL_5$ is systematically negligible.

Resolution of instruments is noted $BL_6$. It is determined with a rectangular distribution by the value read on each voltmeter for power meter. Resolution is then no worse than 5x$10^{-7}$. $BL_6$=5x$10^{-7}/\sqrt{3}$ =2.89x$10^{-7}$.

Contribution of the electronics and especially the use of automatic/manual range is noted $BL_7$. We deduce from knowledge of experimenter that this influence is no more than 2.5x$10^{-3}$. $BL_7$=0.0025/$\sqrt{3}$ =1.44x$10^{-3}$.

Contribution of vibrations due to the environment is noted $BL_8$. We do not operate in case of known vibration source in the environment and the table is enough robust to prevent diffusion of vibration. Pneumatic legs support the optical table. The spectrometer can then operate, but it is only isolated against building vibrations and not against vibrations introduced directly into the table. We operate in safe conditions and avoid any vibration due to components placed on the table. This term can be negligible in normal operating conditions.

The total contribution of BL=$\Sigma BL_i$ is the arithmetic sum of each elementary contribution. It is determine to be BL=4.91x$10^{-7}$+5.77x$10^{-11}$+3.89x$10^{-3}$+2.89x$10^{-7}$+1.44x$10^{-3}$=5.33x$10^{-3}$.

4.3. Estimation of the global uncertainty

Uncertainty at a 1 σ interval of confidence is calculated as follows:

(v) $\quad\quad\quad\quad u_c = \sqrt{(A^2 + BR^2 + BL^2)}$

We deduce from equation (v) that uncertainty at 1 sigma, noted $u_c$, is better than $\sqrt{[(2.89\times10^{-4})^2+0^2+(5.33\times10^{-3})^2]}$. Its leads to a global uncertainty of ±5.34x$10^{-3}$ at 1 σ.

For convenience, and to keep an operational uncertainty in case of degradation or drift of any elementary term of uncertainty, it is wise to slightly degrade the global uncertainty. This is why we choose to keep U = ±1.20x$10^{-2}$ at 2 σ for a common use, corresponding to a voluntary degradation of the uncertainty evaluation assuming that $u_c$ < 6.00x$10^{-3}$. This final uncertainty is defined at 2 σ, according to the empirical rule as 68.27% at 1 σ is not enough, but 95.45% at 2 σ is more efficient for a normal distribution in statistics.

## 5. Conclusion

We first remind that Brillouin Scattering Stimulation of bulk materials and some thin films is non-intrusive and relatively easy to perform. We note that further improvements are still in progress especially for sending light through the sample to be characterized, through its backside. The main result our work, is that after checking the agreement between the measured 15.7 GHz shift on a Brillouin signal on PMMA and the expected value, we make an estimation of the uncertainty on the frequency of the shifted signal corresponding to the speed of phonons inside a material, according to the standards of metrology. The uncertainty is assumed to be of $\pm 1.20 \times 10^{-2}$ at 2 σ for an accurate measurement.

## References

1. Curie J.; Curie P. Développement par compression de l'électricité polaire dans les cristaux hémièdres à faces inclines. *Bulletin de Minéralogie* **1880**, 4-4, 90-93. https://doi.org/10.3406/bulmi.1880.1564

2. Brillouin, L. Diffusion de la lumière et des rayons X par un corps transparent homogène. Influence de l'agitation thermique. *Ann. Phys.* **1922**, 17, 88–122. https://doi.org/10.1051/anphys/192209170088

3. Govind P. Agrawal. Chapter 9 - Stimulated Brillouin scattering. Nonlinear Fiber Optics (Sixth edition) **2019**, 355-399. https://doi.org/10.1016/B978-0-12-817042-7.00016-6

4. Blachowicz T.; Grimsditch M. Scattering, Inelastic: Brillouin. *Encyclopedia of Condensed Matter Physics* **2005**, 199-205. https://doi.org/10.1016/B0-12-369401-9/00643-4

5. Mandelstam L. Light scattering by inhomogeneous media. *Zh. Russ. Fiz. Khim. Ova.* **1926**, 58, 381.

6. Kargar F.; Balandin A.A. Advances in Brillouin–Mandelstam light-scattering spectroscopy. *Nat. Photon.* **2021**, 15, 720–731. https://doi.org/10.1038/s41566-021-00836-5

7. Raman, C. V.; Krishnan, K. S. A new type of secondary radiation. *Nature* **1928**, 121, 501–502. https://doi.org/10.1038/121501C0

8. Kojima, S. 100th Anniversary of Brillouin Scattering: Impact on Materials Science. *Materials* **2022**, 15, 3518. https://doi.org/10.3390/ma15103518

9. Grimsditch M. H.; Ramdas A. K. Brillouin scattering in diamond. *Phys. Rev. B* **1975**, 11, 3139. https://doi.org/10.1103/PhysRevB.11.3139

10. Sandercock, J. Brillouin scattering study of SbSI using a doubled-passed stabilised scanning interferometer. Opt. Commun. **1970**, 2, 73–76. https://doi.org/10.1016/0030-4018(70)90047-7

11. Lindsay S. M.; Anderson M. W.; Sandercock J. R. Construction and alignment of a high performance multipass Vernier tandem Fabry–Perot interferometer. *Review of Scientific Instruments* **1981**, 52(10), 1478-1486. https://doi.org/10.1063/1.1136479.

12. Dil J. G.; van Hijningen N. C. J. A.; van Dorst F.; Aarts R. M. Tandem multipass Fabry-Perot interferometer for Brillouin scattering. *Applied Optics* **1981**, 20(8), 1374-1381. https://doi.org/10.1364/AO.20.001374

13. Hillebrands B. Progress in multipass tandem Fabry-Perot interferometry: I. A fully automated, easy to use, self-aligning spectrometer with increased stability and flexibility. *Review of Scientific Instruments* **1999**, 70(3), 1589-1598. https://doi.org/10.1063/1.1149637

14. Scarponi F.; Mattana S.; Corezzi S.; Caponi S.; Comez L.; Sassi P.; Morresi A.; Paolantoni M.; Urbanelli L.; Emiliani C.; Roscini L.; Corte L.; Cardinali G.; Palombo F.; Sandercock J. R.; Fioretto D. High-Performance Versatile Setup for Simultaneous Brillouin-Raman Microspectroscopy. *Physical Review X* **2017**, 7, 031015. https://doi.org/10.1103/PhysRevX.7.031015

15. GUM: Guide to the Expression of Uncertainty in Measurement, fundamental reference document, JCGM100:2008 (GUM 1995 minor corrections): https://www.bipm.org/en/committees/jc/jcgm/publications (accessed 27 February 2023)

16. Thomas von Clarmann; Steven Compernolle; Frank Hase. Truth and uncertainty. A critical discussion of the error concept versus the uncertainty concept. *Atmospheric Measurement Techniques* **2022**, 15, 1145-1157. https://doi.org/10.5194/amt-15-1145-2022

17. Jong Wha Lee; Euijin Hwang; Raghu N. Kacker. True value, error, and measurement uncertainty: two views. *Accreditation and Quality Assurance* **2022**, 27, 235-242. https://doi.org/10.1007/s00769-022-01508-9

18. Kacker, R.; Sommer, K. D.; Kessel, R. Evolution of modern approaches to express uncertainty in measurement. *Metrologia* **2007**, 44(6)513–529. https://doi.org/10.1088/0026-1394/44/6/011

19. Salzenstein, P.; Pavlyuchenko, E.; Hmima, A.; Cholley, N.; Zarubin, M.; Galliou, S.; Chembo, Y. K.; Larger, L. Estimation of the uncertainty for a phase noise optoelectronic metrology system. *Physica Scripta* **2012**, T149, 014025. https://doi.org/10.1088/0031-8949/2012/T149/014025

20. Pavlyuchenko, E.; Salzenstein, P. Application of modern method of calculating uncertainty to microwaves and opto-electronics. Laser Optics, 2014 International Conference, Saint Petersburg, Russia, June 30 2014-July 4 **2014**, 6886449. https://doi.org/10.1109/LO.2014.6886449

21. Salzenstein, P.; Pavlyuchenko, E. Uncertainty Evaluation on a 10.52 GHz (5 dBm) Optoelectronic Oscillator Phase Noise Performance. *Micromachines* **2021**, 12, 474. https://doi.org/10.3390/mi12050474

22. Won-Kyu Lee; Dai-Hyuk Yu; Chang Yong Park; Jongchul Mun. The uncertainty associated with the weighted mean frequency of a phase-stabilized signal with white phase noise. *Metrologia* **2010**, 47(1), 24–32. https://doi.org/10.1088/0026-1394/47/1/004

23. Salzenstein, P.; Wu, T. Y. Uncertainty analysis for a phase-detector based phase noise measurement system. *Measurement* **2016**, 85, 118–123. https://doi.org/10.1016/j.measurement.2016.02.026

24. Gough, W. The graphical analysis of a Lorentzian function and a differentiated Lorentzian function. *Journal of Physics A: General Physics* **1968**, 1, 704. https://doi.org/10.1088/0305-4470/1/6/309

25. Salzenstein P.; Kuna A.; Sojdr L.; Chauvin J. Significant step in ultra high stability quartz crystal oscillators. *Electronics Letters* **2010**, 46(21), 1433–1434, (2010). https://doi.org/10.1049/el.2010.1828

26. Salzenstein P.; Cholley N.; Kuna A.; Abbé P.; Lardet-Vieudrin F.; Sojdr L.; Chauvin J. Distributed amplified ultra-stable signal quartz oscillator based. *Measurement* **2012**, 45(7), 1937–1939. https://doi.org/10.1016/j.measurement.2012.03.035

27. Howard, L; Stone, J; Fu, J. Real-time displacement measurements with a Fabry-Perot cavity and a diode laser. *Precision Engineering* **2001**, 25(4), 321-335. https://doi.org/10.1016/S0141-6359(01)00086-1

28. Joo, K.-N.; Ellis, J. D.; Spronck, J. W.; Munnig Schmidt, R. H. Design of a folded, multi-pass Fabry–Perot cavity for displacement metrology. *Measurement Science and Technology* **2009**, 20, 107001. https://doi.org/10.1088/0957-0233/20/10/107001

29. Köning, R.; Flügge, J.; Bosse, H. A method for the in situ determination of Abbe errors and

their correction. *Measurement Science and Technology* **2007**, *18*, 476. https://doi.org/10.1088/0957-0233/18/2/S21

30. Leach, R. Abbe Error/Offset. *CIRP Encyclopedia of Production Engineering* **2014**, 1–4. https://doi.org/10.1007/978-3-642-35950-7_16793-1